%% file: main_EB2.tex
\documentclass[aps,prd,nofootinbib,twocolumn,10pt,superscriptaddress]{revtex4-2}
\usepackage{color,amsmath,amssymb,graphicx,latexsym,lineno}
\usepackage{threeparttable,multirow,txfonts,subfigure,booktabs}

\usepackage{accents,slashed,ulem}

\newlength{\dhatheight}

\graphicspath{{./}{Figs_unbin/}}

\begin{document}

\title{Constraints on Lorentz Invariance Violation from PeV Gamma Rays of Cygnus X-3 Observed by LHAASO}

\input{ApJ20260713}

\date{\today}

\begin{abstract}
The Large High Altitude Air Shower Observatory (LHAASO) has detected photons with energies up to 3.7 PeV from the binary system Cygnus X-3 during a flaring episode. The combination of such extreme photon energies and the large source distance provides a sensitive probe of Lorentz invariance violation (LIV). In superluminal LIV scenarios, high-energy photons can become unstable and rapidly decay, resulting in a sharp spectral cutoff. However, the observed hard intrinsic spectrum of Cygnus X-3 exhibits no evidence for such a feature. This absence enables us to place stringent constraints on LIV, with the second-order LIV energy scale constrained to be above $10^{-2}M_{\rm pl}$, improving previous limits by nearly an order of magnitude.

\end{abstract}

\maketitle

\section{introduction}

Lorentz invariance (LI) is a cornerstone of special relativity and a fundamental symmetry of modern particle physics and quantum field theory. However, various approaches to quantum gravity and physics beyond the Standard Model suggest that LI may be violated at energy scales approaching the Planck scale $M_{\rm pl}$ \cite{QG_S:1,QG_S:2,QG_S:3,QG_S:4,QG_S:5,QG_S:6,QG_S:7,QG_S:8}. While such effects are expected to be highly suppressed at accessible energies, high-energy astrophysical observations provide a sensitive probe due to the extreme energies and long propagation distances involved.

High-energy gamma-ray observations have been widely used to test Lorentz invariance violation (LIV) through energy-dependent time delays, threshold effects, and polarization-dependent phenomena \cite{Pulsar_LIV_2017,GRB_fermi,Vasileiou:2013vra,ma_2018,LHAASO:2024lub,AGN_Abdalla_2019,AGN_LIV_Abramowski_2011,Galaverni_2008_Cherenkov1,vac_birefringence_10.1093/mnras/stw2007,vac_birefringence_10.1093/mnras/stz594,Satunin:2019gsl,PhysRevD.95.063001,CANGAROO:Crab,Albert:2019nnn}. In superluminal LIV scenarios, high-energy photons become unstable and can undergo photon decay into electron–positron pairs or photon splitting into multiple lower-energy photons. This induces strong energy-dependent attenuation and leads to a sharp cutoff in the gamma-ray spectrum.

A commonly used description of LIV is provided by modified dispersion relations (MDRs), which parametrize deviations from the standard energy–momentum relation. For photons, the MDR can be written as \cite{MDR1_Coleman_1999,MDR2_Amelino_Camelia_1998,MDR3_AmelinoCamelia:2001dy,MDR4_Ahluwalia_1999}
\begin{equation}
E_\gamma^2 - p_\gamma^2 = \pm |\alpha_n| p_\gamma^{n+2},
\label{mdr}
\end{equation}
where the $\pm$ sign corresponds to superluminal and subluminal cases, respectively. The parameter $\alpha_n$ characterizes the $n$-th order LIV effect and is related to the LIV energy scale via $E_{\rm LIV}^{(n)} = \alpha_n^{-1/n}$.

In the superluminal case, the photon decay process $\gamma \rightarrow e^+e^-$ occurs once the kinematic threshold for electron–positron pair production is exceeded, leading to strong suppression of high-energy photons. 
{If the astrophysical $\gamma$-ray spectrum extends up to $E_\gamma$ without a cutoff,} this implies the following constraints on the LIV energy scale for $n=1$ and $n=2$:

\begin{equation}
\begin{aligned}
E_{\mathrm{LIV}}^{(1)} &\gtrsim 9.57 \times 10^{23} \, \mathrm{eV}\left(\frac{E_\gamma}{\mathrm{TeV}}\right)^3 \\
E_{\mathrm{LIV}}^{(2)} &\gtrsim 9.78 \times 10^{17} \, \mathrm{eV}\left(\frac{E_\gamma}{\mathrm{TeV}}\right)^2\ \ \ .
\end{aligned}
\label{eq:LIV_electron}
\end{equation}

Superluminal LIV can also induce photon splitting processes, $\gamma \rightarrow N\gamma$, with the dominant channel being $\gamma \rightarrow 3\gamma$ \cite{Astapov:2019xmt,Rubtsov:2016bea}. For $n=2$, the decay rate is
\begin{equation}
\Gamma_{\gamma \rightarrow 3\gamma}
=5 \times 10^{-14}
\frac{E_\gamma^{19}}{m_e^8 E_{\mathrm{LIV}}^{(2)10}}\ \ .
\label{eq:gamma}
\end{equation}
{The decay length is then given approximately by $L\sim 1/\Gamma_{\gamma \rightarrow 3\gamma}$. After propagation over an astrophysical distance $L$, the strong energy dependence of the decay rate also leads to a sharp cut-off in the $\gamma$-ray spectrum \cite{Astapov:2019xmt,Satunin:2019gsl,Rubtsov:2016bea}. The non-observation of such a cutoff leads to the following constraint on LIV energy scale as}

\begin{equation}
E_{\mathrm{LIV}}^{(2)}>3.33 \times 10^{19} \mathrm{eV}
\left(\frac{L}{\mathrm{kpc}}\right)^{0.1}
\left(\frac{E_\gamma}{\mathrm{TeV}}\right)^{1.9}.
\label{eq:LIV_3gamma}
\end{equation}

Recently, the LHAASO collaboration reported the detection of gamma rays from the binary system Cygnus X-3, located at a distance of approximately $9~\mathrm{kpc}$ \cite{Reid:2023ksq}. LHAASO has detected photons from this source with energies extending up to 3.7 PeV \cite{LHAASO:2025ysm}. The observation of multi-PeV photons from such a distant source provides a sensitive probe of 
possible LIV effects.
In this work, we use the LHAASO observations of Cygnus X-3 to test superluminal LIV and derive limits on the LIV energy scale from the survival of multi-PeV photons over astrophysical distances.

\section{The LHAASO experiment and data set}

The Large High Altitude Air Shower Observatory (LHAASO) is a ground-based cosmic-ray and gamma-ray observatory located at an altitude of 4410 m in Daocheng, Sichuan Province, China \cite{LHAASO:2019qtb}. It consists of three sub-arrays: the Kilometer Square Array (KM2A), the Water Cherenkov Detector Array (WCDA), and the Wide Field-of-view Cherenkov Telescope Array (WFCTA). Among them, KM2A provides the primary sensitivity to gamma rays above $\sim 10$ TeV. The KM2A array consists of 5216 electromagnetic detectors (EDs) and 1188 muon detectors (MDs), covering an area of approximately 1.3 km$^2$. The arrival direction and shower core position are reconstructed using the timing and charge information from the EDs. The shower energy is estimated using $\rho_{50}$, defined as the particle density at 50 m from the shower axis and obtained by fitting the modified NKG lateral distribution function \cite{LHAASO:KM2A_perf}. The energy resolution improves from about $17\%$ at 0.1 PeV to $10\%$ at 1 PeV. The cosmic-ray background is efficiently suppressed by exploiting the enhanced muon content in hadronic showers, achieving a rejection power of $\sim 1.0 \times 10^{4}$ at $\sim 100$ TeV and increasing further at higher energies \cite{2024ChPhC}.

During the operation period from December 26, 2019 to July 31, 2024, four flaring episodes of Cygnus X-3 were recorded by KM2A. These correspond to the Modified Julian Date (MJD) intervals 58921.7–59179.7, 59275.7–59419.7, 59473.7–59569.7, and 60289.7–60523.0. After applying the same event selection criteria adopted in previous work\cite{LHAASO:2025ysm}, a total of 692.8 days of flaring-period data remain.

To search for possible LIV-induced spectral features, both binned and unbinned analyses are performed. In the binned analysis, the sky is divided into $0.1^\circ \times 0.1^\circ$ pixels in equatorial coordinates, and events are accumulated accordingly. The energy range from 10 TeV to 1 PeV is divided into bins of $\Delta\log_{10}(E_{\rm{rec}})=0.2$, while a finer binning of $\Delta\log_{10}(E_{\rm{rec}})=0.1$ is adopted above 1 PeV. The background is estimated using the direct integration method, taking advantage of events with similar directions but different arrival times in the local coordinate system\cite{Fleysher_2004}.
At energies above $\sim 1$ PeV, where statistics are extremely limited, the background is stabilized by scaling the detector acceptance measured at $> 100$ TeV. This procedure has a negligible impact on the systematic uncertainty due to the very low background rate in this energy regime. Details of the unbinned analysis and alternative background treatments are provided in Sec. \ref{sec:ubinned_method} of the Supplemental Material.

\begin{figure}
	\centering
    \includegraphics[width=8.6cm]{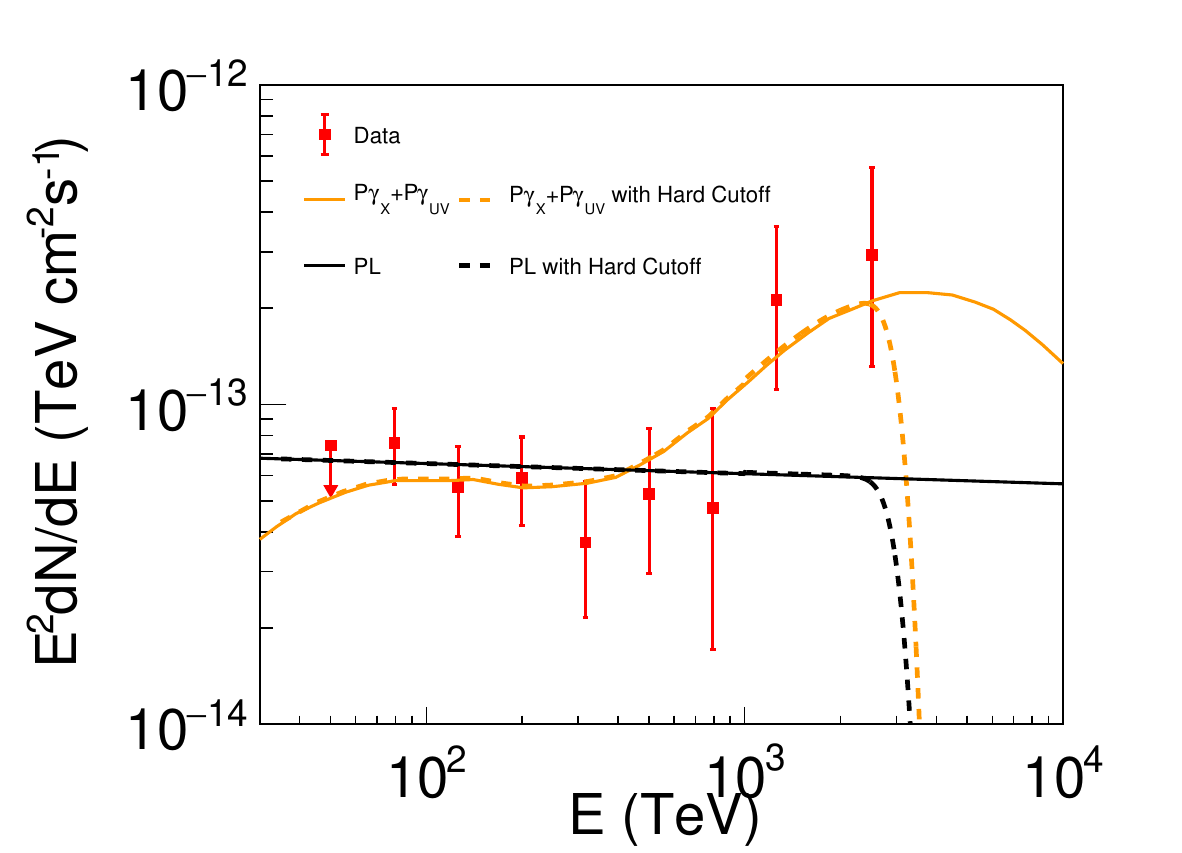}
	\caption{{Comparison between the best-fit intrinsic spectrum of Cygnus X-3 (solid line) and a hypothetical model with a hard cutoff at 3 PeV (dashed line). The latter has been convolved with the instrument's energy resolution.} }	\label{fig:SED}
\end{figure}

\section{Method}

In projection on the sky, Cygnus X-3 lies near the center of the Cygnus Bubble \cite{Cygnus_Bubble}, a region containing multiple gamma-ray sources. To disentangle possible source confusion, a three-dimensional (spatial–spectral) fitting framework is adopted. In this analysis, the morphological and spectral parameters of nearby sources are fixed to the values determined in Ref.~\cite{LHAASO:2025ysm}. 

At a distance of $\sim9$ kpc, Cygnus X-3 is treated as a point source. After correcting for the absorption of photons by the interstellar radiation field (ISRF) and the cosmic microwave background (CMB), the intrinsic spectrum can be described by a photomeson production scenario in the innermost region of the relativistic jet, where protons are accelerated to tens of PeV. We adopt this $P\gamma_X+P\gamma_{UV}$ spectral model as our baseline for studying LIV effects (Fig.~\ref{fig:SED}); detailed calculations are provided in Sec. \ref{sec:Photomeson} in the Supplemental Material. To examine the dependence on the intrinsic spectral assumption, a power-law (PL) model is also tested, as shown by the black solid line in Fig.~\ref{fig:SED}. Both $\gamma \to e^+ e^-$ and $\gamma \to 3\gamma$ processes predict spectra with hard cutoffs (dashed lines in Fig.~\ref{fig:SED}), which can be parametrized as
\begin{equation}
f(E)=f_{0}(E) H(E_{\rm{cut}} - E)\ \, ,
\label{equ:sed1}
\end{equation}
where $E$ represents the true energy, $f_{0}(E)$ represents the intrinsic spectrum, $E_{\rm cut}$ denotes the cutoff energy and $H(E_{\rm cut}-E)$ is the Heaviside step function used to model the LIV effect.

A profile likelihood test is used to evaluate the compatibility of the spectrum with a LIV-induced hard cutoff. In the binned analysis, the likelihood function is defined as

\begin{equation}
\mathcal{L}(E_{\rm cut}) =
\prod_{i,j} \mathrm{Poisson}\left(N_{\rm obs}^{i,j},
N_{\rm sig}^{i,j}(E_{\rm cut}) + N_{\rm bkg}^{i,j}\right),
\end{equation}
where $i$ and $j$ denote the energy and spatial bins, respectively.
$N_{\rm obs}^{i,j}$ is the observed number of events, $N_{\rm bkg}^{i,j}$ is the estimated background contribution, and
$N_{\rm sig}^{i,j}$ is the expected signal obtained by folding the source model with the detector response.

\begin{figure}
	\centering
    \includegraphics[width=8.6cm]{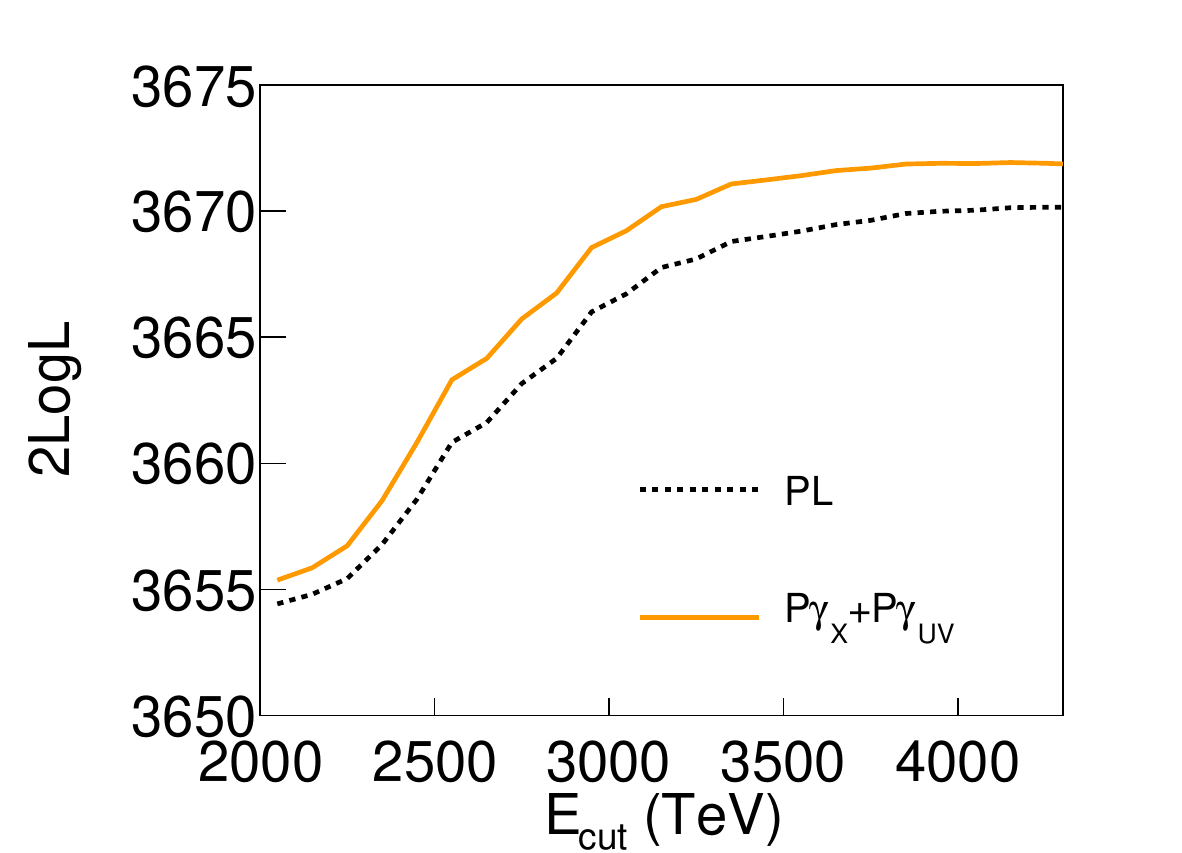}
	\caption{The profile log-likelihood as a function of $E_{\rm cut}$ for Cygnus X-3. The solid (dotted) curve corresponds to the $P\gamma_X+P\gamma_{UV}$ (PL) spectral model. }
	\label{fig:2LogL}
\end{figure}

The profile likelihood as a function of $E_{\rm cut}$ is shown in
Fig.~\ref{fig:2LogL}. The likelihood increases with increasing cutoff energy, indicating that the data favor the absence of a spectral cutoff at the highest energies. We therefore define the test statistic (TS)  to determine the lower limit on the cutoff
energy:
\begin{equation}
\mathrm{TS}(E_{\rm cut}) = 2\ln \frac{\mathcal{L}(E_{\rm cut}\to\infty)}{\mathcal{L}(E_{\rm cut})}.
\end{equation}
The null hypothesis corresponds to no cutoff ($E_{\rm cut}\to\infty$), while finite $E_{\rm cut}$ represents the LIV scenario. 

Because of the limited statistics in the highest-energy bins, the TS distribution does not follow the $\chi^2$ distribution \cite{Algeri2020_wilk}. Hence, we adopt the $CLs$ method \cite{Read_2002,LHAASO:2021opi} to derive the limit of $E_{\rm cut}$. The TS distributions under both hypotheses are therefore obtained from Monte Carlo (MC) simulations. For the null hypothesis, simulated datasets are generated using the expected signal with $E_{\rm cut}\rightarrow\infty$ together with the measured background. For each tested $E_{\rm cut}$ value, the corresponding alternative distributions are obtained in the same way. The resulting TS distributions are used to construct the $CL_s$ value,

\begin{equation}
CL_s=\frac{CL_{s+b}}{CL_b},
\end{equation}
where $CL_{s+b}$ and $CL_b$ are the probabilities of obtaining a TS value larger than the observed one under the alternative and null hypotheses, respectively.

{For the unbinned approach, the likelihood function is defined as
\begin{equation}
\mathcal{L}_{unbin}(E_{\rm cut})=\frac{(n^{\rm{exp}}_{\rm{sig}}+n^{\rm{exp}}_{\rm{bkg}})^{N}}{N!}e^{-(n^{\rm{exp}}_{\rm{sig}}+n^{\rm{exp}}_{\rm{bkg}})} \prod_{i=1}^{N} f(E_{\rm cut}; \alpha,\delta, E_{\mathrm{rec}}),
\end{equation}
where $N$ denotes the total number of events within the region of interest (ROI), and $n^{\rm exp}_{\rm sig}$ and $n^{\rm exp}_{\rm bkg}$ denote the expected numbers of signal and background events, respectively. The function $f(E_{\rm cut}; \alpha,\delta, E_{\mathrm{rec}})$ represents the probability density of observing an event with right ascension $\alpha$, declination $\delta$, and reconstructed energy $E_{{\rm rec}}$. Details of the probability density used in the unbinned fit are provided in Sec. \ref{sec:ubinned_likelihood} in the Supplemental Material.}

\section{Results}

\begin{figure}
	\centering
    \includegraphics[width=8.6cm]{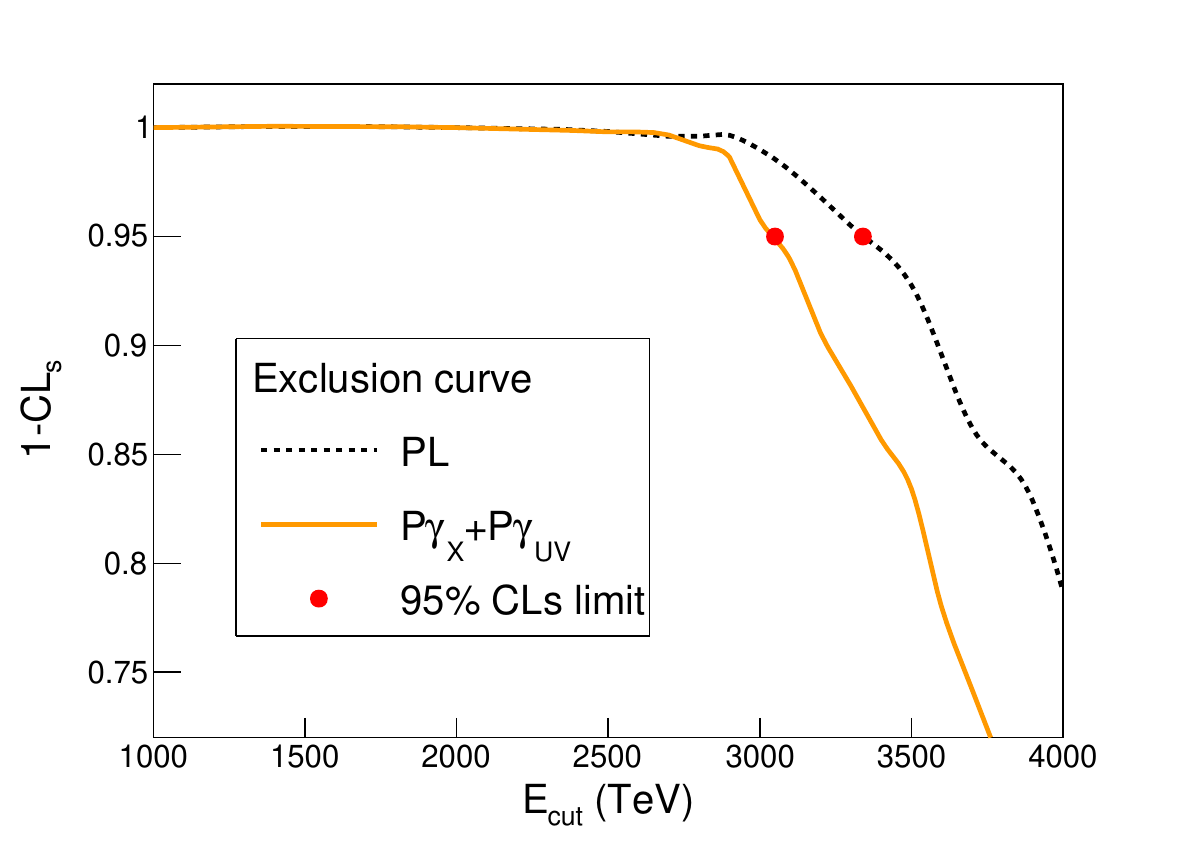}
	\caption{The probability $1-CL_s$ as a function of $E_{\rm cut}$ for different spectral models. The red dots mark the $E_{\rm cut}$ values which are excluded at the 95\% CL with the $CL_s$ method. 
	}     \label{fig:cls_ecut}
\end{figure}

In the binned analysis, the TS distributions are derived for each tested value of $E_{\rm cut}$ under both the null (LI) and alternative (LIV) hypotheses. The corresponding $1-CL_s$ values as a function of $E_{\rm cut}$ are shown in Fig.~\ref{fig:cls_ecut}. For the baseline $P\gamma_X+P\gamma_{UV}$ spectral model, the lower limit on the cutoff energy at the $95\%$ confidence level is $E_{\rm cut}=3.05~\mathrm{PeV}$ (the corresponding TS distributions are shown in Sec. \ref{sec:TS_distributions} in  Supplemental Material). The PL model gives a slightly higher limit because it predicts fewer events in the $\log_{10}(E_{\rm{rec}})=3.4$--3.6 energy range, where two events are observed (See Sec. \ref{sec:data_CX3} of the Supplemental Material), thus providing stronger constraints on low-energy cutoff scenarios. The results obtained with the PL spectral model are treated as a systematic uncertainty.

The constraints on the LIV energy scale are derived from $E_{\rm cut}$ using Eqs.~\ref{eq:LIV_electron} and \ref{eq:LIV_3gamma} for the $\gamma\rightarrow e^+e^-$ and $\gamma\rightarrow3\gamma$ processes. Table~\ref{tab:Ec_res} summarizes the resulting $95\%$ confidence-level limits on $E_{\rm cut}$ and the corresponding LIV energy scales. For first-order LIV, the derived limit reaches $\sim2.72\times10^{34}~\mathrm{eV}$, far above the Planck scale ($M_{\rm pl}\approx1.22\times10^{28}~\mathrm{eV}$), providing a stringent constraint on this scenario. For second-order LIV, the photon decay process gives a lower limit of $E_{\rm LIV}^{(2)}\sim9.09\times10^{24}~\mathrm{eV}$, while the photon splitting process yields $E_{\rm LIV}^{(2)}\sim1.73\times10^{26}~\mathrm{eV}$. The latter constraint depends only weakly on the source distance and reaches the \(10^{-2}M_{\rm pl}\) level.

As a consistency check, we perform an unbinned analysis using the same \(CL_s\) procedure and obtain \(E_{\rm cut}^{95\%}=3.06~\mathrm{PeV}\), consistent with the baseline binned result (Fig.~\ref{fig:fig5}). This agreement demonstrates the robustness of our constraint and confirms that the adopted binning scheme does not significantly affect the sensitivity. Details of the unbinned analysis are provided in Sec. II in the Supplemental Material.

\begin{figure}
	\centering
	\includegraphics[width=8.6cm]{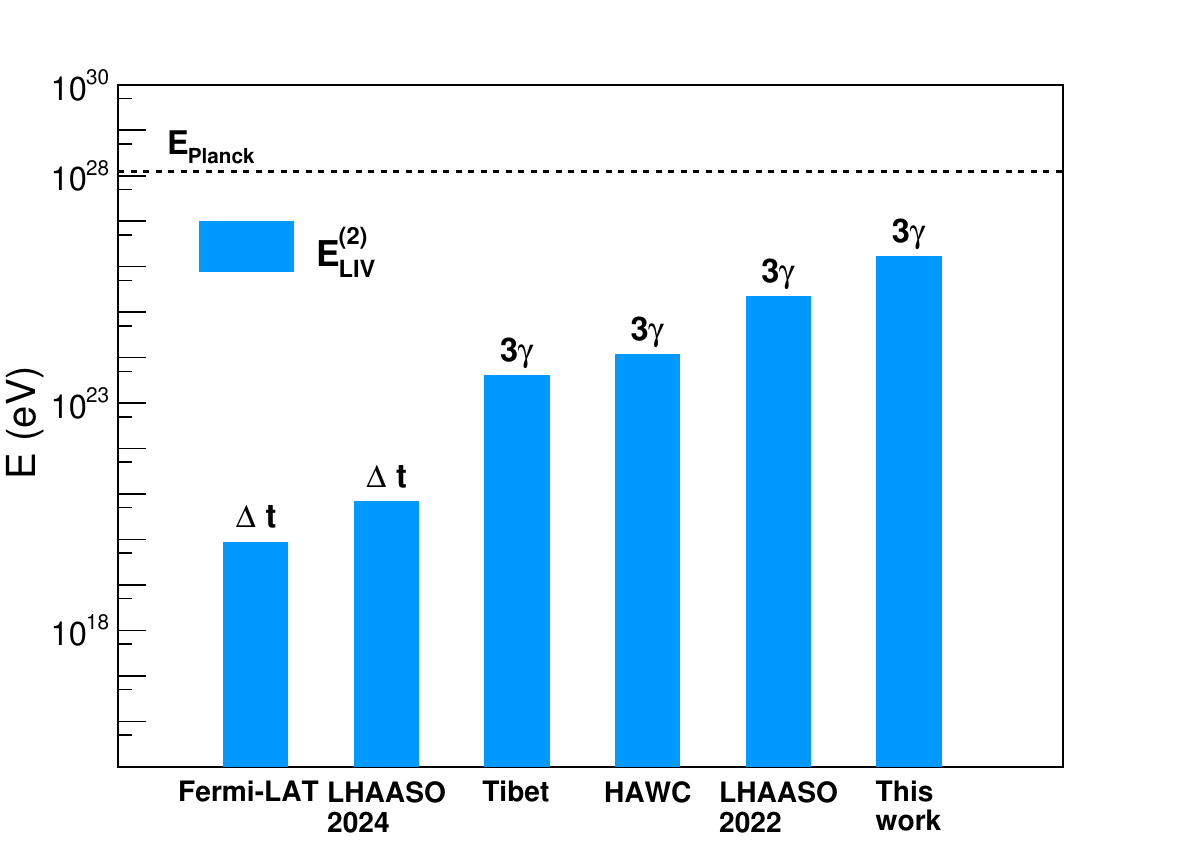}
	\caption{
	{Comparison of the constraints on $E^{(2)}_{\rm LIV}$ derived from LHAASO and other experiments \cite{Satunin:2019gsl,Albert:2019nnn,LHAASO:2021opi,Vasileiou:2013vra,LHAASO:2024lub}. Here, $3\gamma$ denotes the photon-splitting process, while $\Delta t$ denotes the time-delay method. }}
	\label{fig:comp}
\end{figure}

A comparison with previous constraints is shown in Fig.~\ref{fig:comp}. We include limits on photon decay and photon splitting from Tibet \cite{Satunin:2019gsl}, HAWC \cite{Albert:2019nnn}, and the previous LHAASO measurement \cite{LHAASO:2021opi}, as well as constraints from energy-dependent 
time-delay constraints derived from Fermi-LAT observations of gamma-ray bursts \cite{Vasileiou:2013vra} and from LHAASO observations of GRB 221009A \cite{LHAASO:2024lub}. Our results improve previous constraints by nearly one order of magnitude and provide one of the most stringent limits on second-order superluminal LIV. Since second-order LIV is CPT-invariant and cannot be effectively constrained by vacuum birefringence measurements, our results provide an important complementary constraint on this class of LIV models.

\begin{table*}[htbp]
	\caption{Source distance \(L\), maximum reconstructed photon energy \(E_{\rm rec}^{\max}\), $95\%$ CL lower limit on the cutoff energy \(E_{\rm cut}^{95\%}\), and the corresponding lower limits on the first- and second-order LIV energy scales derived from the \(\gamma\rightarrow e^+e^-\) and \(\gamma\rightarrow3\gamma\) processes. The quoted uncertainties are systematic. } 
	\centering 
	\renewcommand\tabcolsep{10.0pt}

	\begin{tabular}{c ccccccccc} 
		\hline\hline 
		Source & $L$ & $E_{\rm{rec}}^{\rm max}$ & $E_{\rm cut}^{95\%}$ & $E_{\rm LIV}^{(1)}$ & $E_{\rm LIV}^{(2)}$ & $E_{\rm LIV}^{(2)}$ $(3\gamma)$\\
		& (kpc) & (PeV) & (PeV) &  (eV) & (eV) & (eV) \\ 
		& & & & $\times 10^{32}$ & $\times 10^{23}$ &  $\times 10^{25}$ \\ [0.5ex]
		\hline      
		Cygnus X-3 & 8.95 & $3.735 \pm 0.411 $ & $3.05^{+0.29}_{-0.29}$ & $271.9^{+85.2}_{-70.4}$ & $90.9^{+18.1}_{-16.5}$ & $17.3^{+3.3}_{-3.0}$ \\ 
		\hline
		\hline 
	\end{tabular}
	\label{tab:Ec_res}
\end{table*}

\section{Systematic uncertainties}

Several sources of systematic uncertainty affect the derived LIV energy-scale constraints. The dominant contribution arises from the uncertainty in the intrinsic spectrum of Cygnus X-3. By comparing the baseline $P\gamma_X+P\gamma_{UV}$ model with the PL spectral model, we find a difference of $\sim9.5\%$ in $E_{\rm cut}^{95\%}$. 

The uncertainty from the background estimation is negligible, as the analysis is nearly background-free above PeV energies. Additional uncertainties arise from atmospheric simulations, which introduce uncertainties of approximately $7\%$ and 0.02 in the flux and spectral index, respectively. These effects propagate to a $\sim1.5\%$ uncertainty in $E_{\rm cut}^{95\%}$.

Combining all contributions in quadrature gives a total systematic uncertainty of $\sim9.6\%$ in $E_{\rm cut}^{95\%}$ and the corresponding LIV energy scales are reported in Table~\ref{tab:Ec_res}.

\section{Summary}

In this study, we constrain Lorentz invariance violation using recent LHAASO observations of the binary system Cygnus X-3. These observations revealed photons with energies up to 3.7 PeV during several flaring episodes. The detection of multi-PeV photons from a distant source ($\sim9$ kpc) provides a powerful test of photon stability under superluminal LIV scenarios. Such scenarios predict energy-dependent attenuation of high-energy photons through processes including photon decay and splitting, which would modify the observed spectrum with a cutoff-like feature. However, no evidence for such attenuation is found in the observed hard spectrum. We therefore derive lower limits on the LIV energy scale. For second-order LIV, which is CPT-invariant, our results constrain the energy scale to be above $10^{-2}M_{\rm pl}$, improving previous limits by nearly one order of magnitude and providing one of the most stringent constraints in this channel.

\section{Acknowledgements}
We would like to thank the staff members of National Facility for Large High Altitude Air Shower Observatory (LHAASO) (CSTR: \href{https://cstr.cn/31117.02.LHAASO}{31117.02.LHAASO.ED and 31117.02.LHAASO.MD}) who work at the LHAASO site above 4400 meter above the sea level year round to maintain the detector and keep the water recycling system, electricity power supply and other components of the experiment operating smoothly. We are grateful to Chengdu Management Committee of Tianfu New Area for the constant financial support for research with LHAASO data. We appreciate the computing and data service support provided by the National High Energy Physics Data Center for the data analysis in this paper. This research work is supported by the following grants: National Key R\&D Program of China under grants 2025YFE0202600, The National Natural Science Foundation of China No.12522510, No.12447105, No.12393851, No.12393852, No.12393853, No.12393854, NSFC No.12261160362, No.12305120, No.12375107, and in Thailand from the NSRF via the Research and Innovation Acceleration Agency for Competitiveness and Area Development (RCAD) (Program Management Unit for Technology and Innovation for Future Industries (PMU-B) : Brainpower for Future Industries) [grant number B39G690003].

\newpage
\bibliographystyle{apsrev}
\bibliography{refs}

\clearpage

\input{LIV_SM}

\end{document}

%% file: ApJ20260713.tex
\author{Zhen Cao}
\affiliation{State Key Laboratory of Particle Astrophysics \& Experimental Physics Division \& Computing Center, Institute of High Energy Physics, Chinese Academy of Sciences, 100049 Beijing, China}
\affiliation{University of Chinese Academy of Sciences, 100049 Beijing, China}
\affiliation{Tianfu Cosmic Ray Research Center, 610000 Chengdu, Sichuan,  China}
 
\author{F. Aharonian}
\affiliation{Tianfu Cosmic Ray Research Center, 610000 Chengdu, Sichuan,  China}
\affiliation{University of Science and Technology of China, 230026 Hefei, Anhui, China}
\affiliation{Yerevan State University, 1 Alek Manukyan Street, Yerevan 0025, Armenia }
\affiliation{Max-Planck-Institut for Nuclear Physics, P.O. Box 103980, 69029  Heidelberg, Germany}
 
\author{Y.X. Bai}
\affiliation{State Key Laboratory of Particle Astrophysics \& Experimental Physics Division \& Computing Center, Institute of High Energy Physics, Chinese Academy of Sciences, 100049 Beijing, China}
\affiliation{Tianfu Cosmic Ray Research Center, 610000 Chengdu, Sichuan,  China}
 
\author{Y.W. Bao}
\affiliation{Tsung-Dao Lee Institute \& School of Physics and Astronomy, Shanghai Jiao Tong University, 200240 Shanghai, China}
 
\author{D. Bastieri}
\affiliation{Center for Astrophysics, Guangzhou University, 510006 Guangzhou, Guangdong, China}
 
\author{X.J. Bi}
\affiliation{State Key Laboratory of Particle Astrophysics \& Experimental Physics Division \& Computing Center, Institute of High Energy Physics, Chinese Academy of Sciences, 100049 Beijing, China}
\affiliation{University of Chinese Academy of Sciences, 100049 Beijing, China}
\affiliation{Tianfu Cosmic Ray Research Center, 610000 Chengdu, Sichuan,  China}
 
\author{Y.J. Bi}
\affiliation{State Key Laboratory of Particle Astrophysics \& Experimental Physics Division \& Computing Center, Institute of High Energy Physics, Chinese Academy of Sciences, 100049 Beijing, China}
\affiliation{Tianfu Cosmic Ray Research Center, 610000 Chengdu, Sichuan,  China}
 
\author{W. Bian}
\affiliation{Tsung-Dao Lee Institute \& School of Physics and Astronomy, Shanghai Jiao Tong University, 200240 Shanghai, China}
 
\author{J. Blunier}
\affiliation{APC, Universit'e Paris Cit'e, CNRS/IN2P3, CEA/IRFU, Observatoire de Paris, 119 75205 Paris, France}
 
\author{A.V. Butkevich}
\affiliation{Institute for Nuclear Research of Russian Academy of Sciences, 117312 Moscow, Russia}
 
% ...
\collaboration{The LHAASO Collaboration}

\author{H.B. Jiang}
\affiliation{State Key Laboratory of Particle Astrophysics \& Experimental Physics Division \& Computing Center, Institute of High Energy Physics, Chinese Academy of Sciences, 100049 Beijing, China}
\affiliation{Tianfu Cosmic Ray Research Center, 610000 Chengdu, Sichuan,  China}
\affiliation{School of Physics and Microelectronics, Zhengzhou University, 450001 Zhengzhou, Henan, People’s Republic of China}

%% file: LIV_SM.tex
\linenumbers

\graphicspath{{Figs/},{Figs_unbin/}}
\onecolumngrid

\setcounter{figure}{0}
\renewcommand\thefigure{S\arabic{figure}}
\setcounter{table}{0}
\setcounter{equation}{0}
\setcounter{section}{0}
\renewcommand\thetable{S\arabic{table}}



\begin{center}
    {\Large Supplemental Material of ``Constraints on Lorentz Invariance Violation from PeV Gamma Rays of Cygnus X-3 Observed by LHAASO''}\\
    (The LHAASO Collaboration)\\
\end{center}

\section{Binned Analysis}

\subsection{Data in the Cygnus X-3 region}
\label{sec:data_CX3}
In this section, we present the event counts within a $0.5^\circ$ region of interest (ROI) centered on Cygnus X-3 for energies above 100~TeV. The data are binned in $\log_{10}(E_{\rm{rec}}/\mathrm{TeV})$, as shown in Table \ref{tab:data_1deg}. 

\begin{table}[!ht]
\centering
\begin{tabular}{c|c|c|c|c|c|c|c|c|c|c|c|c} \hline\hline
$\log_{10}(E_{\rm{rec}})$  & 2.0-2.2 & 2.2-2.4 & 2.4-2.6 & 2.6-2.8 & 2.8-3.0 & 3.0-3.1 & 3.1-3.2 & 3.2-3.3 & 3.3-3.4 & 3.4-3.5 & 3.5-3.6& $>3.6$ \\ \hline
on & 46 & 17 & 7 & 4 & 4 & 1 & 2 & 0 & 0 & 1 & 1 & 0   \\ \hline
bkg & 4.8700 & 1.4519 & 0.5031 & 0.1747 & 0.0593 & 0.0107 & 0.0061 & 0.0037 & 0.0021 & 0.0009 & 0.0003 & 0.0006  \\ \hline
Ns (others) &  24.2699 & 5.0839 & 1.3618 & 0.5140 & 0.2097 & 0.0542 & 0.0303 & 0.0206 & 0.0127 & 0.0075 & 0.0051 & 0.0063   \\ \hline
Ns ($P\gamma_{X}+P\gamma_{UV}$) &  18.4885 & 11.9145 & 7.9810 & 5.1808 & 3.2535 & 1.1762 & 0.8424 & 0.7282 & 0.5848 & 0.4728 & 0.4296 & 0.9108  \\ \hline
Ns (PL) &  20.2558 & 13.1907 & 8.4872 & 4.7480 & 2.1675 & 0.5683 & 0.3320 & 0.2408 & 0.1717 & 0.1261 & 0.1100 & 0.2626   \\ 
\hline \hline
\end{tabular}
\caption{Event counts within a $0.5^\circ$ ROI centered on Cygnus X-3 in different energy bins. 
``on'' denotes the number of observed events, ``bkg'' represents the estimated cosmic-ray background, and Ns (others) is the expected gamma-ray contribution from other sources in the region excluding Cygnus X-3. Ns ($P\gamma_{X}+P\gamma_{UV}$) and Ns (PL) denote the expected gamma-ray contributions from Cygnus X-3, assuming $P\gamma_{X}+P\gamma_{UV}$ and PL spectral shapes without a hard cutoff, respectively.}
\label{tab:data_1deg}
\end{table}

\subsection{TS distributions for $E_{cut}^{95\%}$}
\label{sec:TS_distributions}
The probability distribution of the TS is obtained via MC simulations for a given $E_{\rm cut}$. Under the null hypothesis, the expected signal $N_{\rm sig}$ is computed from the  intrinsic spectrum assuming $E_{\rm cut}\to\infty$, and is convolved with the detector response. Mock observed counts $N_{\rm on}$ are generated as Poisson realizations of $N_{\rm sig}+N_{\rm bkg}$, where $N_{\rm bkg}$ is estimated from data. The TS values are then calculated for each realization, yielding the TS distribution under the null hypothesis. For the alternative hypothesis, TS distributions are obtained in the same way by generating mock data with finite $E_{\rm cut}$ for each tested value. We define $CL_{s+b}$ ($CL_b$) as the probability of obtaining a TS value greater than that observed under the alternative (null) hypothesis. The modified confidence level is then defined as $CL_s = CL_{s+b}/CL_b$. A given $E_{\rm cut}$ is excluded at $95\%$ confidence level if $CL_s = 0.05$.

\begin{figure}
	\centering
	\includegraphics[width=8.6cm]{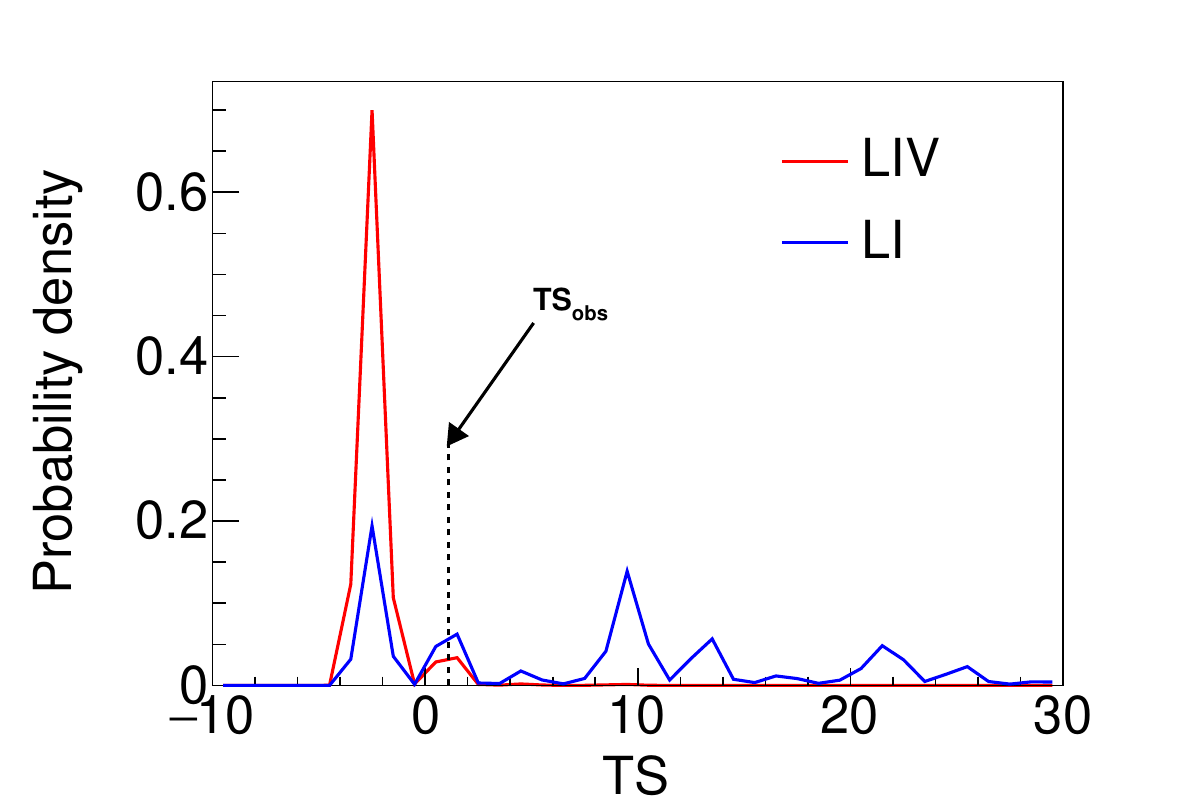}
	\caption{TS distributions of MC data with and without hard cutoff assuming the $P\gamma_{X}+P\gamma_{UV}$ spectral model and $E_{\rm cut} = 3050$ TeV. The red and blue lines are fitted results based on 50,000 MC simulations for the LIV and LI cases, respectively. }
	\label{fig:TS_dis}
\end{figure}

Fig.~\ref{fig:TS_dis} shows the TS distributions for $E_{\rm cut}=3.05~\mathrm{PeV}$ assuming the $P\gamma_{X}+P\gamma_{UV}$ spectral model. The MC simulation is repeated $5\times10^{4}$ times. The red and blue histograms correspond to the LIV and LI hypotheses, respectively, while the black dashed line indicates the observed TS. The TS distributions exhibit nontrivial features due to limited statistics in the high-energy bins (see Table~\ref{tab:data_1deg}). In the LIV scenario with $E_{\rm cut}=3.05~\mathrm{PeV}$, the expected signal above the cutoff is strongly suppressed, and most MC realizations contain zero events in these bins. This leads to a dominant peak and typically yields $\mathcal{L}_0 < \mathcal{L}_1$. In realizations with nonzero counts, the likelihood ordering reverses ($\mathcal{L}_0 > \mathcal{L}_1$), producing  structures at ${\rm TS} > 0$. A similar behavior is observed for the LI case, where fluctuations between zero and nonzero counts in the high-energy bins generate features on both sides of ${\rm TS}=0$. 

\section{Unbinned Analysis}
\label{sec:ubinned_method}
\subsection{Data and unbinned maximum-likelihood method}
\label{sec:ubinned_likelihood}
\begin{figure}[h]
	\centering
	\includegraphics[width=4.0in,height=2.0in]{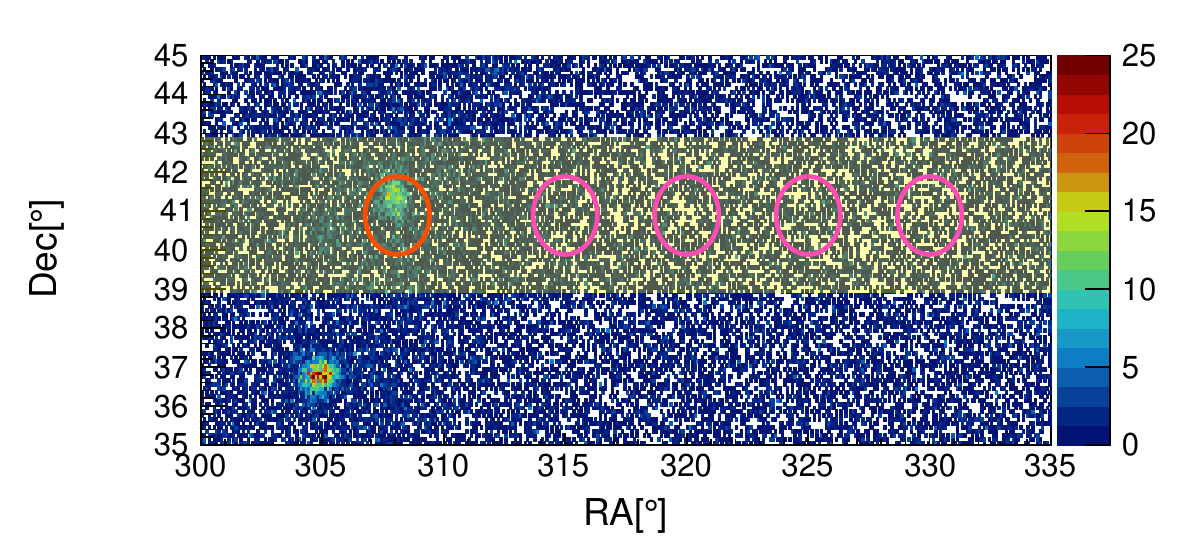}
	\caption{
The distribution of gamma-like events with energies above 40 TeV in the equatorial coordinate system (right ascension and declination) around Cygnus X-3.  The sky map is divided into $0^{\circ}.1 \times 0^{\circ}.1$  square pixels, according to their arrival direction. The red circle is the signal ROI, while the four magenta circles denote the background regions used for background estimation within the signal ROI. 
}
	\label{fig:fig1}
\end{figure}

The unbinned maximum-likelihood method is expected to yield more precise measurements than the binned method in the low-statistics regime, as binning inevitably leads to information loss and consequently larger statistical uncertainties in the parameter estimates.
A previous study of Cygnus X-3~\cite{LHAASO:2025ysm} indicates that the statistics in the source region become very limited above several hundred TeV.
To cross-check and improve the result derived from the binned analysis, we develop a three-dimensional unbinned likelihood fitting method, incorporating two spatial dimensions and energy, to extract the second-order LIV energy scale. The same data set and selection criteria as in the binned analysis are adopted in this work. After event selection, Fig.~\ref{fig:fig1} shows the sky map for gamma-like events with energy above 40 TeV. 
We define an ROI as a $1^{\circ}$-radius circle centered on Cygnus X-3, with location taken from a previous measurement~\cite{LHAASO:2025ysm}, to perform an unbinned maximum-likelihood fit. The likelihood function is defined as
\begin{equation}
\mathcal{L}_{unbin}=\frac{(n^{\rm{exp}}_{\rm{sig}}+n^{\rm{exp}}_{\rm{bkg}})^{N}}{N!}e^{-(n^{\rm{exp}}_{\rm{sig}}+n^{\rm{exp}}_{\rm{bkg}})} \prod_{i=1}^{N} f(\theta,\phi,t, E_{\mathrm{rec}}),
\end{equation}
where $N$ denotes the total number of events within the ROI, and $n^{\rm exp}_{\rm sig}$ and $n^{\rm exp}_{\rm bkg}$ are the expected numbers of gamma-ray signal and cosmic-ray background events within the ROI. The function $f(\theta, \phi, t, E_{\mathrm{rec}})$ represents the probability of detecting an event with zenith $\theta$ and azimuth $\phi$ in the horizon coordinate system, reconstructed energy $E_{\mathrm{rec}}$, at time $t$.


The probability density function (PDF) $f(\theta,\phi,t,E_{\mathrm{rec}})$ is given by 
\begin{equation}
    f(\theta,\phi,t, E_{\mathrm{rec}})=f_s(\theta,\phi,t, E_{\mathrm{rec}})+\beta f_b(\theta,\phi,t,E_{\rm rec}),
\end{equation}
where $f_s(\theta,\phi, t, E_{\rm rec})$ and $f_b(\theta,\phi, t, E_{\rm rec})$ are the PDFs of the gamma-ray signal and the cosmic-ray background, respectively, and $\beta$ is the background normalization relative to the signal. The hard cutoff in the energy spectrum at $E_{\rm cut}$ induced by the LIV effect can be modeled using the Heaviside step function, $H(E_{\rm cut} - E)$.
The PDF for the signal component consists of the source morphology and the spectral energy distribution (SED), and  can be calculated as follows:
\begin{equation}
\begin{aligned}
    f_s(\theta,\phi, t, E_{\rm rec}, E_{\rm cut}) = &\sum_{i} \Big[ 
    f_{\rm Mor}^i(\alpha_i, \delta_i) \otimes \mathrm{PSF}(\theta, E_{\rm rec})\\
    &\times \int f_{\rm SED}^{i}(E_{\rm truth})H(E_{\rm cut}-E_{\rm truth}) S_{0} \cos(\theta)
    \, \eta(\theta, E_{\rm rec}, E_{\rm truth}) dE_{\rm truth},
\Big]
\end{aligned}
\label{eq:3}
\end{equation}
where $f_{\rm Mor}^i(\alpha_i, \delta_i)$ represents the spatial morphology function of gamma-ray source $i$, with $\alpha_i$ and $\delta_i$ denoting the corresponding right ascension and declination.
${\rm PSF}(\theta, E_{\rm rec})$ is the point spread function, $f_{\rm SED}^i(E_{\rm truth})$ is the spectral energy distribution for source i, and $S_{0}$ is the total detector area used in the MC simulation. 
{Here $\eta(\theta,E_{\rm rec}, E_{\rm truth})$ denotes the KM2A detection efficiency evaluated using MC data \cite{LHAASO:2024zug}.} $E_{\rm truth}$represents the true energy of gamma-rays in the MC simulation.

The use of Eq.~\ref{eq:3} to compute $f_s(\theta,\phi, t, E_{\rm rec})$ involves a large number of coordinate transformations between the horizon and equatorial coordinate systems, leading to a substantially increased computational cost. To reduce the cost, the following simplifications are adopted:
\begin{itemize}
    \item  
    Given that the gamma-ray sources at the same declination share the same zenith-angle distribution, Eq.~\ref{eq:3} can be simplified to depend only on the $(\alpha,\delta)$ of each event. As a result, $f_s(\theta,\phi,t,E_{\rm rec})$ is reduced to $f_s(\alpha,\delta,E_{\rm rec})$. The corresponding PSF can be transformed into $\mathrm{PSF}(\delta,E_{\rm rec})$ by incorporating the zenith-angle distribution appropriate to each declination.
    
    \item Since this work focuses exclusively on the LIV effect, we assume that the spectral energy distribution of Cygnus X-3 is identical across all four flare periods.
\end{itemize}

Hence, Eq.~\ref{eq:3} can be simplified to 
\begin{equation}
\begin{aligned}
    f_s(\alpha,\delta, E_{\rm rec}, E_{cut}) = &\sum_{i} \Big[ 
    f_{\rm Mor}^i(\alpha_i, \delta_i) \otimes \mathrm{PSF}(\delta_i, E_{\rm rec})\\
   & \times \int\int f_{\rm SED}^i(E_{\rm truth})H( E_{\rm cut}-E_{\rm truth}) S_0 \cos(\theta)
    \, \eta(\theta, E_{\rm rec}, E_{\rm truth}) \, T \, f(\theta)_{\delta_i} \, d\theta \, dE_{\rm truth}
\Big].
\end{aligned}
\end{equation}
where $\rm{PSF}(\delta_i,E_{rec})$ is the average PSF calculated according to the corresponding zenith-angle distribution $f(\theta)_{\delta_i}$ over the observation period $T$. Due to the limited MC statistics, we calculate the $\eta(\theta,E_{\rm rec}, E_{\rm truth})$ in bins: $\Delta (\log_{10}(E_{\rm rec}/{\rm TeV}))=0.05$ over the range of 4 TeV to 4 PeV, $\Delta (\log_{10}(E_{\rm truth}/{\rm TeV}))=0.08$ in the range of 1 TeV - 10 PeV, and $\Delta \theta =1^{\circ}$ for zenith angles less than $50^{\circ}$.  
The response matrix precision is limited by MC statistics, which in turn leads to a non-negligible systematic uncertainty. In the future, this issue will be addressed by increasing the MC statistics, thereby enabling a more precise result from the unbinned fit.
Linear interpolation is applied between adjacent $E_{\rm rec}$ bins to calculate  $\eta(\theta,E_{\rm rec}, E_{\rm truth})$.

For the background PDF, a model-independent, data-driven approach is employed to determine the cosmic-ray background distribution in the energy spectrum.  Due to the very low statistics above several hundred TeV, we use the energy distribution of events from a declination band twice as wide as the signal ROI (the yellow band in Fig.~\ref{fig:fig1}) to model the cosmic-ray background spectrum within the signal ROI. The corresponding number of background events is estimated from  four source-free magenta circles in Fig.~\ref{fig:fig1} (denoted as background regions), at the same declination as the signal ROI. 
The spatial distribution of background is assumed to be uniform within this small signal ROI. This assumption is validated in the background regions by a fit yielding $\chi^2/\rm ndf = 1.2$.


\subsection{Fit Results}

In the unbinned likelihood fit, we adopt the same spatial and spectral models as those used in the previous study~\cite{LHAASO:2025ysm} for all gamma-ray sources. Except for Cygnus X-3, the parameters of all other sources are fixed to the values obtained from previous measurements~\cite{LHAASO:2025ysm}. 
The location and the normalization of the spectral energy model for Cygnus X-3 are allowed to vary in the fit. The expected number of background events derived from the background regions is allowed to vary under a Gaussian constraint, with the width set by the uncertainty on the estimated number.

\begin{figure}
	\centering
	\includegraphics[width=8.8cm]{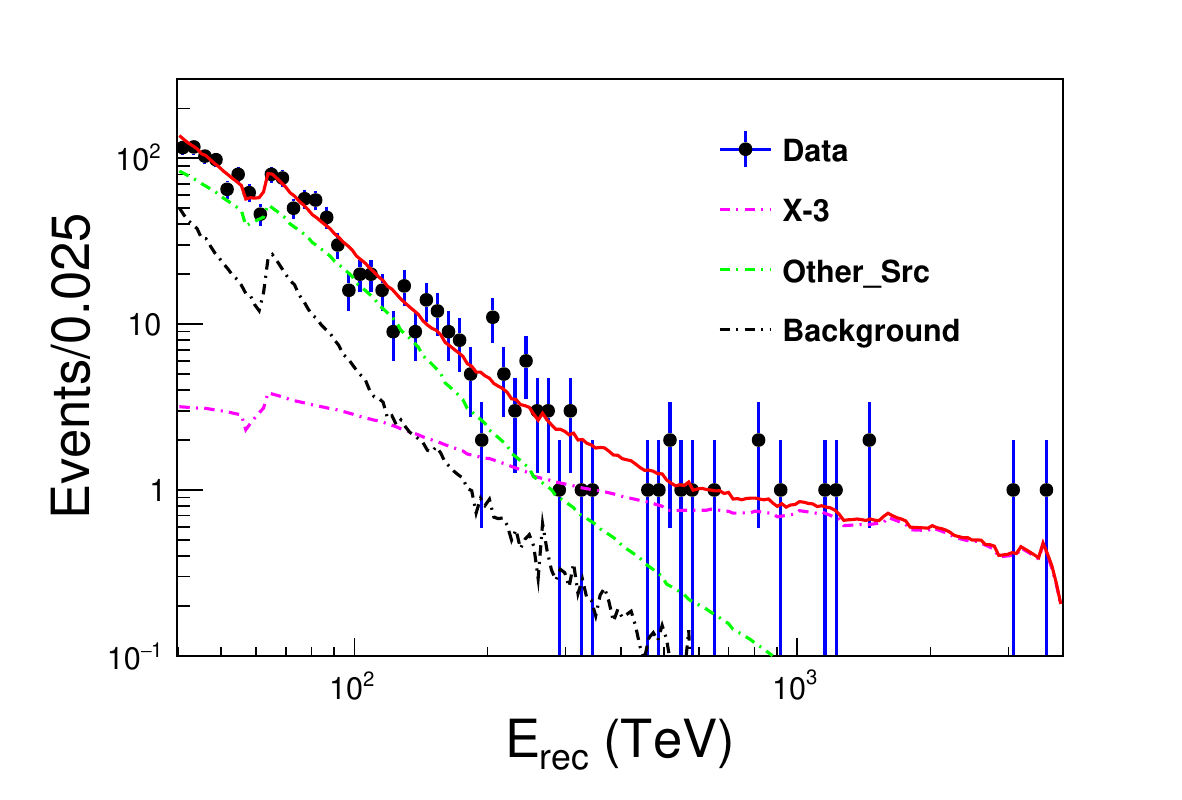}
	\caption{
    Fit result of the spectral energy distribution for events in the signal ROI. The data points are shown with error bars. The red solid curve is the total fit. The magenta dashed curve corresponds to the fitted contribution from Cygnus X-3, the green dashed curve represents contributions from other sources, and the black dashed curve indicates the background contribution. The feature around $(E_{\rm rec}) \sim 60$~TeV arises from a significant change in the gamma–proton discrimination parameter.
    The fluctuation observed above 1 PeV is due to the limited MC statistics.
    }
	\label{fig:fig2}
\end{figure}

\begin{figure}[htpb]
	\centering
	\includegraphics[width=8.6cm]{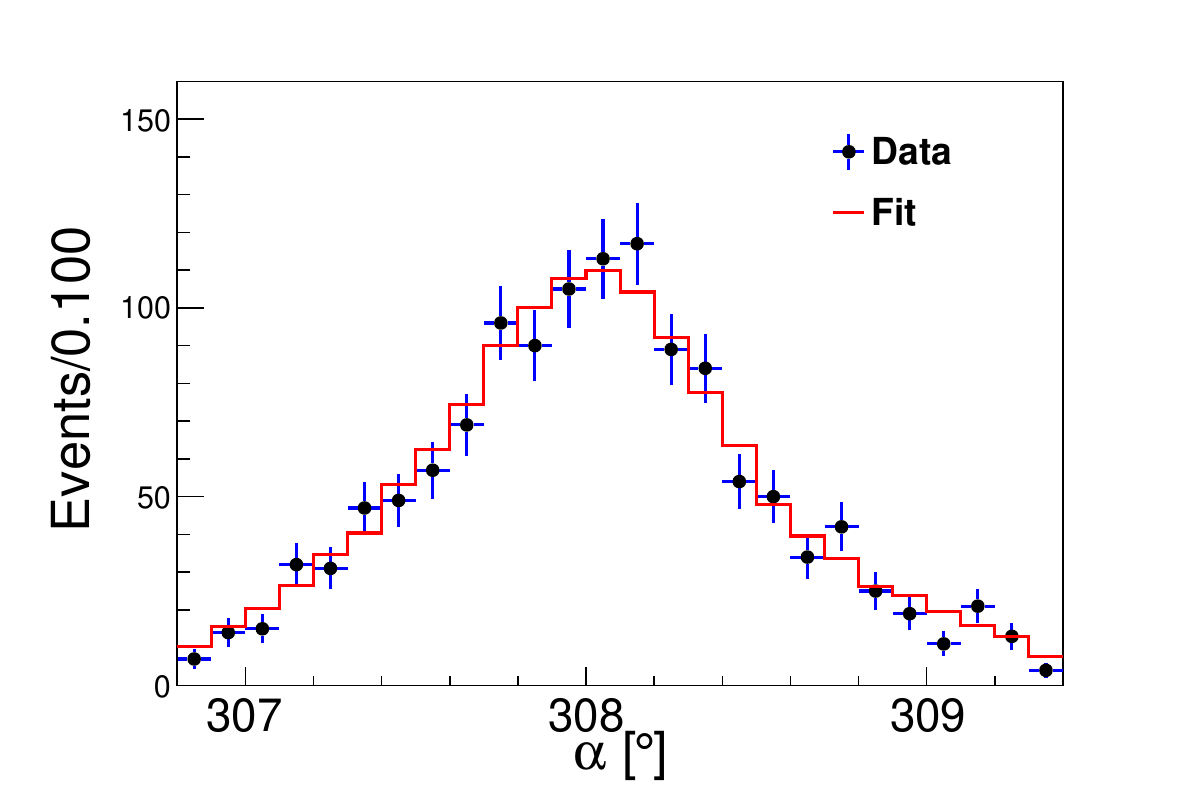}
 	\includegraphics[width=8.6cm]{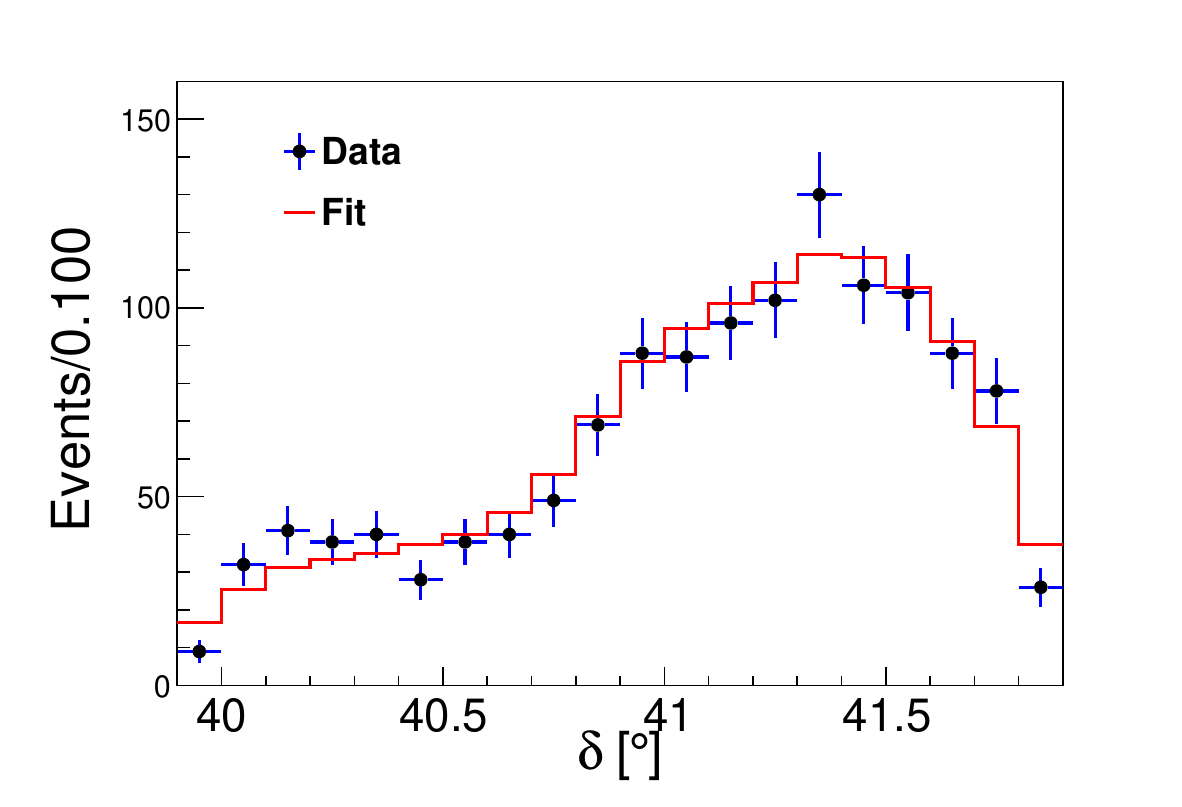}
	\caption{ The projections of the fit results on right ascension and declination. The dots with error bars are data, and the red solid curve is the total fit. }
	\label{fig:fig3}
\end{figure}

The projections of the fit results onto energy spectrum, right ascension and declination are shown in Fig.~\ref{fig:fig2} and Fig.~\ref{fig:fig3}.
The best-fit coordinates are $\alpha = 308.09^{\circ} \pm 0.03^{\circ}$ and $\delta = 40.95^{\circ} \pm 0.03^{\circ}$, and the spectral normalization of $F_0 = 0.94 \pm 0.12$; all of which are consistent with the results reported in the previous study~\cite{LHAASO:2025ysm}. 
The best-fit cutoff energy is $\log_{10}(E_{\rm cut}/{\rm TeV})=3.75$, with a statistical significance below $1\sigma$. Therefore, the data show no evidence for a LIV-induced spectral cutoff.

\subsection{$1-CL_s$ distribution}

\begin{figure}[htpb]
	\centering
	\includegraphics[width=8.6cm]{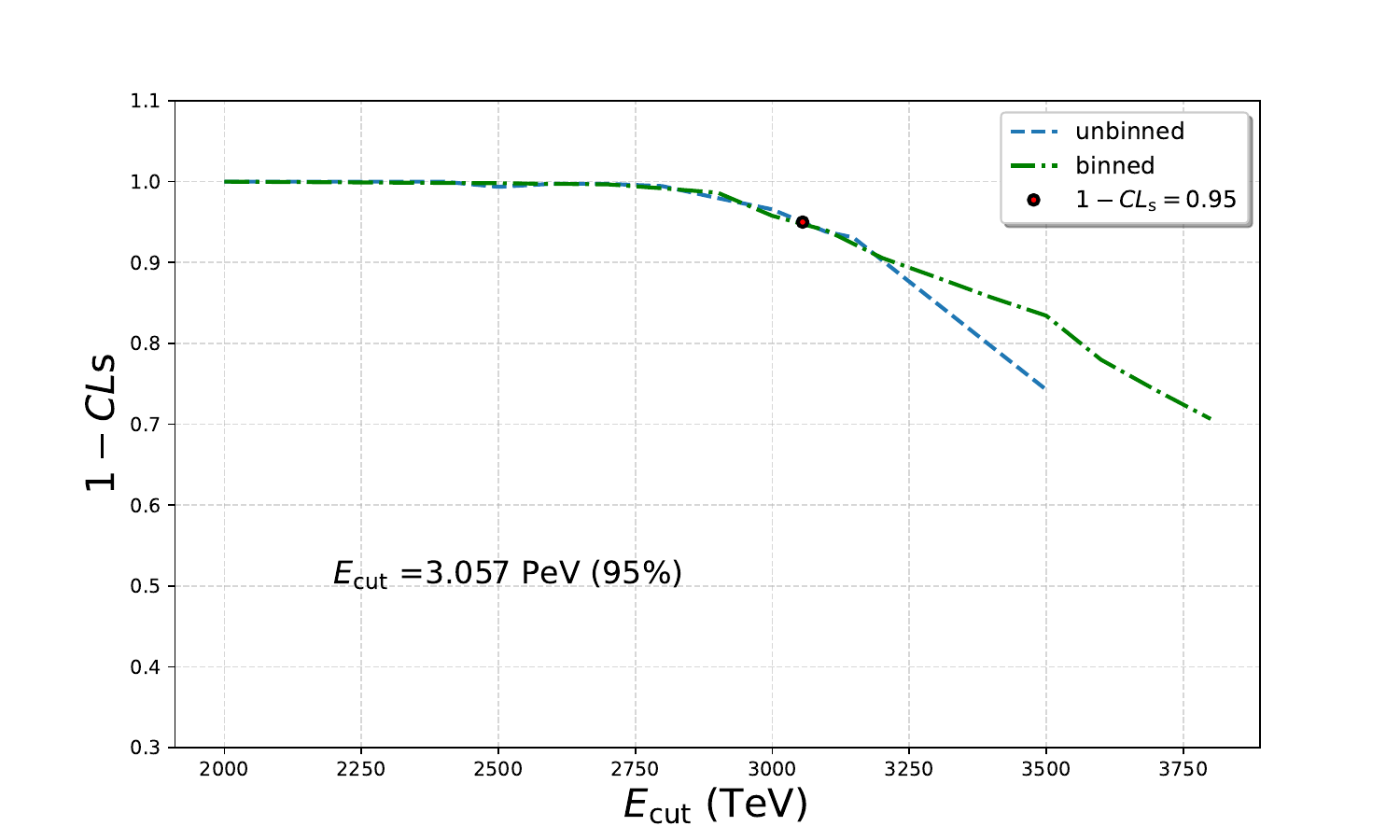}
	\caption{Comparison of $1 - CL_s$ distributions from unbinned and binned analyses for the $P\gamma_X + P\gamma_{UV}$ spectral model.}
	\label{fig:fig5}
\end{figure}

We also apply the \(CL_{\rm s}\) method to the unbinned analysis as an independent cross-check. The corresponding \(1-CL_{\rm s}\) distribution is shown in Fig.~\ref{fig:fig5}. The resulting $95\%$ confidence-level lower limit on the cutoff energy is \(3.057~\mathrm{PeV}\), consistent with the \(3.05~\mathrm{PeV}\) obtained from the baseline binned analysis. This agreement indicates that the adopted binning scheme causes negligible sensitivity loss for the present data set. In the relevant energy range, the bin half-width corresponding to \(\Delta\log_{10}(E_{\rm rec})=0.1\) is approximately \(13\%\), comparable to the detector energy resolution of \(\sim10\%\). Therefore, retaining event-by-event energy information provides limited additional sensitivity.

The unbinned approach further requires a precise characterization of the event-level detector response. The available MC samples \cite{LHAASO:2024zug}, generated with a power-law spectrum of index \(-2.0\), have limited statistics in the PeV range, leading to additional uncertainty in the response determination. By varying the response matrix within its MC statistical uncertainties, we estimate an approximately \(4\%\) uncertainty in the Cygnus X-3 flux above PeV. Therefore, we adopt the binned analysis as the baseline result.

\section{Calculation of the $p\gamma$ Photohadronic Process and $\gamma\gamma$ Absorption}
\label{app:pgamma_calculation}

\subsection{Photomeson Process of Cygnus X-3}
\label{sec:Photomeson}
Protons can be accelerated to tens of PeV in the innermost region of the relativistic jet of Cygnus X-3. The escaping protons then interact with ultraviolet photons ($\epsilon \sim 10\text{ eV}$) from the companion star via the photohadronic ($p\gamma$) process, naturally giving rise to PeV gamma rays.
As detailed in the companion observational study \cite{LHAASO:2025ysm}, an additional component is needed to account for the lower-energy gamma rays. This extra channel could be attributed either to $p\gamma$ interactions between protons and X-ray photons from the accretion disk or corona, or to $pp$ interactions. {In this work, we adopt the $p\gamma_{X}$ process to describe the lower-energy gamma rays; however, this choice is not critical to our results, as the LIV effect is primarily sensitive to high-energy gamma rays.}


The fitting results reported in Ref.~\cite{LhaasoCollaboration:2025jwd} are adopted in this work. The primary accelerated proton spectrum is modeled as a power law with an exponential cutoff, $N_p(E_p) = N_0 E_p^{-\alpha} \exp(-E_p/E_{p,\rm max})$ , with the cutoff energy set to 50 PeV and the spectral index fixed at 2.3. The interaction angle between the proton and the UV photons is assumed to be 79$^{\circ}$.


\subsection{Intrinsic Gamma-Ray Spectrum and $\gamma\gamma$ Absorption}

The TeV-PeV photons are absorbed by lower-energy photon fields through $\gamma$$\gamma$ pair-production. The observed gamma-ray flux $f_{\rm obs}(E_\gamma)$ after absorption is given by:
\begin{equation}
f_{\rm obs}(E_\gamma) = f_0(E_\gamma)\exp\left[-\tau_{\gamma\gamma}(E_\gamma)\right],
\end{equation}
where $f_0(E_\gamma)$ denotes the intrinsic gamma-ray spectrum and $\exp[-\tau_{\gamma\gamma}(E_\gamma)]$ is the attenuation factor due to pair production.

Cygnus X-3 is located in the Galactic plane at a distance of $\sim 9$ kpc. Gamma rays from this source are subject to absorption via $\gamma\gamma$ interactions with photons of the interstellar radiation field (ISRF) and the cosmic microwave background (CMB). The ISRF dominates the optical depth in the energy range $\sim 10$–$200$ TeV, while at higher energies ($\gtrsim 200$ TeV) absorption by the CMB becomes dominant. The corresponding attenuation factor has been calculated in Ref.~\cite{LHAASO:2025ysm} and is adopted in this work. This \(\gamma\gamma\) absorption is included when modeling the intrinsic spectrum.